\documentclass[sigconf]{acmart}
\usepackage{booktabs}
\usepackage{times}
\usepackage{epsfig,endnotes}
\usepackage[T1]{fontenc}
\usepackage{listings}
\usepackage{graphicx}
\usepackage{hyperref}
\usepackage{opera}
\usepackage{url}
\usepackage{tabularx}
\usepackage{multirow}
\usepackage{listings}
\usepackage{mathptmx}
\usepackage{footnote}
\usepackage{pgfplots}
\usepackage{xcolor}
\usepackage{url}
\usepackage{xparse}
\usepackage[htt]{hyphenat}

\ExplSyntaxOn
\NewDocumentCommand{\replace}{mmm}
{
	\marian_replace:nnn {#1} {#2} {#3}
}

\tl_new:N \l_marian_input_text_tl

\cs_new_protected:Npn \marian_replace:nnn #1 #2 #3
{
	\tl_set:Nn \l_marian_input_text_tl { #1 }
	\tl_replace_all:Nnn \l_marian_input_text_tl { #2 } { #3 }
	\tl_use:N \l_marian_input_text_tl
}
\ExplSyntaxOff

\let\OldTexttt\texttt
\newcommand{\breakslash}[1]{\replace{#1}{/}{/\allowbreak}}
\newcommand{\breakpoint}[1]{\replace{#1}{.}{.\allowbreak}}
\renewcommand{\texttt}[1]{\OldTexttt{\breakpoint{\breakslash{#1}}}}

\makesavenoteenv{tabular}
\makesavenoteenv{table}

\renewcommand{\textrightarrow}{$\rightarrow$}

\definecolor{codegreen}{rgb}{0,0.6,0}
\definecolor{codegray}{rgb}{0.5,0.5,0.5}
\definecolor{codepurple}{rgb}{0.58,0,0.82}
\definecolor{backcolour}{rgb}{0.95,0.95,0.92}

\lstdefinestyle{mystyle}{
    backgroundcolor=\color{backcolour},
    commentstyle=\color{codegreen},
    keywordstyle=\color{magenta},
    otherkeywords={make, sudo, curl, uname},
    numberstyle=\tiny\color{codegray},
    stringstyle=\color{codepurple},
    basicstyle=\footnotesize,
    breakatwhitespace=false,
    breaklines=true,
    captionpos=b,
    keepspaces=true,
    numbers=left,
    numbersep=5pt,
    showspaces=false,
    showstringspaces=false,
    showtabs=false,
    tabsize=2
}

\colorlet{punct}{red!60!black}
\definecolor{background}{HTML}{EEEEEE}
\definecolor{delim}{RGB}{20,105,176}
\colorlet{numb}{magenta!60!black}

\expandafter\def\expandafter\UrlBreaks\expandafter{\UrlBreaks  \do\a\do\b\do\c\do\d\do\e\do\f\do\g\do\h\do\i\do\j  \do\k\do\l\do\m\do\n\do\o\do\p\do\q\do\r\do\s\do\t  \do\u\do\v\do\w\do\x\do\y\do\z\do\A\do\B\do\C\do\D  \do\E\do\F\do\G\do\H\do\I\do\J\do\K\do\L\do\M\do\N  \do\O\do\P\do\Q\do\R\do\S\do\T\do\U\do\V\do\W\do\X  \do\Y\do\Z}

\lstdefinelanguage{json}{
    literate=
     *{:}{{{\color{punct}{:}}}}{1}
      {,}{{{\color{punct}{,}}}}{1}
      {\{}{{{\color{delim}{\{}}}}{1}
      {\}}{{{\color{delim}{\}}}}}{1}
      {[}{{{\color{delim}{[}}}}{1}
      {]}{{{\color{delim}{]}}}}{1},
}

\lstdefinelanguage{Ini}
{
    columns=fullflexible,
    morecomment=[s][\color{purple}\bfseries]{[}{]},
    morecomment=[l]{\#},
    morecomment=[l]{;},
    commentstyle=\color{gray}\ttfamily,
    morekeywords={},
    otherkeywords={=,:},
    keywordstyle={\color{green}\bfseries}
}

\begin{document}
	
\copyrightyear{2017}
\acmYear{2017}
\setcopyright{acmcopyright}
\acmConference{SoCC '17}{September 24--27, 2017}{Santa Clara, CA,
	USA}\acmPrice{15.00}\acmDOI{10.1145/3127479.3129249}
\acmISBN{978-1-4503-5028-0/17/09}

\title{Practical Whole-System Provenance Capture}

\author{Thomas Pasquier}
\affiliation{	\institution{Harvard University}
	\city{Cambridge}
	\state{USA}
}
\email{tfjmp@seas.harvard.edu}

\author{Xueyuan Han}
\affiliation{	\institution{Harvard University}
	\city{Cambridge}
	\state{USA}
}
\email{hanx@g.harvard.edu}

\author{Mark Goldstein}
\affiliation{	\institution{Harvard University}
	\city{Cambridge}
	\state{USA}
}
\email{markgoldstein@g.harvard.edu}

\author{Thomas Moyer}
\authornote{Work was completed while author was a member of technical staff at MIT Lincoln Laboratory.}
\affiliation{	\institution{UNC Charlotte}
	\city{Charlotte}
	\state{USA}
}
\email{tmoyer2@uncc.edu}

\author{David Eyers}
\affiliation{	\institution{University of Otago}
	\city{Dunedin}
	\state{New Zealand}
}
\email{dme@cs.otago.ac.nz}

\author{Margo Seltzer}
\affiliation{	\institution{Harvard University}
	\city{Cambridge}
	\state{Massachusetts}
}
\email{margo@eecs.harvard.edu}

\author{Jean Bacon}
\affiliation{	\institution{University of Cambridge}
	\city{Cambridge}
	\state{United Kingdom}
}
\email{jean.bacon@cl.cam.ac.uk}

\renewcommand{\shortauthors}{T. Pasquier et al.}

\begin{abstract} 	Data provenance describes how data came to be in its present form.
It includes data sources and the transformations that have been applied to them.
Data provenance has many uses, from forensics and security to aiding the reproducibility of scientific experiments.
We present CamFlow, a whole-system provenance capture mechanism that
integrates easily into a PaaS offering.
While there have been several prior
whole-system provenance systems that captured a comprehensive, systemic and
ubiquitous record of a system's behavior,
none have been widely adopted. They either 
A) impose too much overhead, 
B) are designed for long-outdated kernel releases and are hard to port to
current systems,
C) generate too much data, or
D) are designed for a single system.
CamFlow addresses these shortcoming by: 
1) leveraging the latest kernel design advances to achieve efficiency;
2) using a self-contained, easily maintainable implementation relying on a Linux Security Module, NetFilter, and other existing kernel facilities;
3) providing a mechanism to tailor the captured provenance data to the needs of the application; and
4) making it easy to integrate provenance across 
distributed systems.
The provenance we capture is streamed and consumed by tenant-built auditor applications.
We illustrate the usability of our implementation by describing three such applications: demonstrating compliance with data regulations; performing fault/intrusion detection; and implementing data loss prevention.
We also show how CamFlow can be leveraged to capture meaningful provenance without modifying existing applications.
 \end{abstract}




\maketitle

\section{Introduction} \label{sec:introduction}
Provenance was originally a formal set of documents describing the origin and ownership history of a work of art.
These documents are used to guide the assessment of the authenticity and quality of the item.
In a computing context, \emph{data provenance} represents, in a formal manner, the relationships between data items (\emph{entities}), transformations applied to those items (\emph{activities}), and persons or organisations associated with the data and transformations (\emph{agents}).
It can be understood as the record of the origin and transformations of data within a system~\cite{carata2014primer}.

Data provenance has a wide range of applications, from facilitating the reproduction of scientific results~\cite{goecks2010galaxy} to verifying compliance with legal regulations (see \autoref{sec:motivation}).
For example, in earlier work~\cite{pasquier2016information}, we discussed how provenance can be used to demonstrate compliance with the French data privacy agency guidelines in a cloud-connected smart home system; in other work~\cite{pasquierpucsi2017} we proposed data provenance to audit compliance with privacy policies in the Internet of Things.

\begin{figure}[t]
	\centering
	\includegraphics[width=\columnwidth]{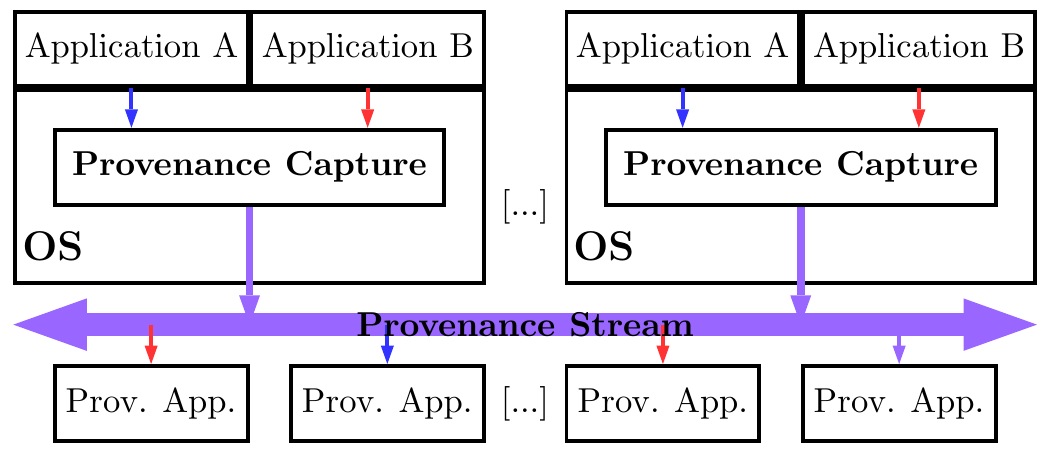}
	\caption{CamFlow cloud provenance architecture.} 
	\label{image:introduction:prov}
\end{figure}

In this paper, we examine the deployment of Provenance as a Service in PaaS clouds, as illustrated in \autoref{image:introduction:prov}.
Tenants' application instances (\eg Docker containers)
run on top of a cloud-provider-managed Operating System (OS).
We include in this OS a provenance capture module that records all interactions between processes and kernel objects (\ie files, sockets, IPC \etc).
We also record network packet information, to track the exchange of data across machines.
We stream the captured provenance data to
tenant-provided \emph{provenance applications} (which can process and/or store the provenance).
We focus on the capture mechanism and discuss four concrete provenance applications in \autoref{sec:examples}.

Two of us were deeply involved in the development of prior, similar
provenance capture systems:
PASS~\cite{muniswamy2006provenance} and LPM~\cite{bates2015trustworthy}.
In both cases, we struggled to keep abreast with current OS releases and
ultimately declared the code unmaintainable.
CamFlow achieves the goals of prior projects, but with a cleaner architecture
that is more maintainable -- we have already upgraded CamFlow several more
times than was ever accomplished with the prior systems.
The adoption of the Linux Security Modelu (LSM) architecture and support for
NetFilters makes this architecturally possible.
Ultimately, our goal is to have provenance integrated into the mainline Linux
kernel.
Thus, we developed a fully self-contained Linux kernel module, discussed in
detail in \autoref{sec:capture}.

A second challenge in whole system provenance capture is the sheer volume of
data, \emph{``which [...] continues to grow indefinitely over time''}~\cite{bates2015trustworthy}.
Recent work~\cite{bates2015take, pasquier2016ic2e} addresses this issue by limiting capture based on security properties, on the assumption that only sensitive objects are of interest.
In practice, after the design stage, many provenance applications answer only a well-defined set of queries.
We extend the ``take what you need'' concept~\cite{bates2015take} beyond the security context,
with detailed policies that capture requirements to meet the exact needs of a provenance application.

The contributions of this work are:
1) a practical implementation of whole-system provenance that 
can easily be maintained and deployed;
2) a policy enforcement mechanism that finely tunes the provenance captured to meet application requirements;
3) an implementation that relies heavily on standards and current practices, interacting with widely used software solutions, designed in a modular fashion to
interoperate with existing software;
4) a new, simpler approach to retrofit provenance into existing ``provenance-unaware'' applications;
5) a demonstration of several provenance applications;
and
6) an extension of provenance capture to containers and shared memory.

\section{The Scope of Provenance} \label{sec:motivation}
Data provenance -- also called data \emph{lineage}, or \emph{pedigree} -- was originally introduced to understand the origin of data in a database~\cite{woodruff1997supporting, buneman2001and}.
Whole-system provenance~\cite{pohly2012hi} goes further, tracking file metadata and transient system objects. It gives a complete picture of a system from \emph{initialisation} to \emph{shutdown}.
We begin by examining a number of illustrative provenance use cases.

\noindgras{Retrospective Security}~\cite{lampson2004computer, povey1999optimistic, weitzner2007beyond, AmirCS2016} is the detection of security violations after execution, by determining whether an execution complied with a set of rules.
We use provenance to detect such violations.
In particular, we verify compliance with contractual or legal regulations~\cite{pasquier2016information}.
We discuss a concrete implementation in \autoref{sec:examples:compliance}.

\noindgras{Reproducibility:}
Many have explored using provenance to guarantee the reproducibility of
computational experiments or to explain the irreproducibility of such results
~\cite{greenwood2003provenance, simmhan2005survey, miles2007requirements}.
It is increasingly common for such experiments to run on a PaaS platform.
Creating a detailed record of the data ((\eg input dataset, parameters \etc)) and software used (\eg libraries, environment \etc) and the dependencies between
them enables reproduction, by creation of
an environment as close as possible to that of the original execution.
Differential provenance techniques~\cite{chen2016good} can provide information to explain why two executions produce different results, which might otherwise
be difficult to understand, \eg in a long and complex scientific workflow.

\begin{figure}[t]
	\centering
	\includegraphics[width=\columnwidth]{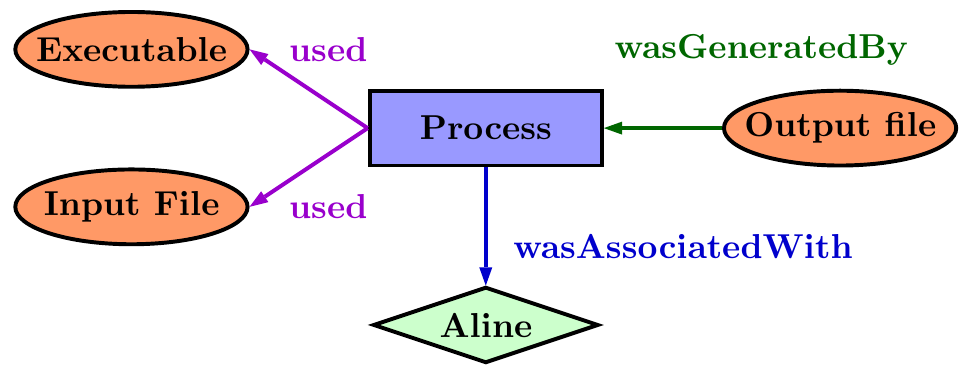}
	\caption{A W3C ProvDM compliant provenance graph.} 
	\label{image:introduction:w3c}
\end{figure}

\noindgras{Deletion and Policy Modification:}
When dealing with highly sensitive data 
it is important to be able to delete every document containing certain information, including 
derived documents.  
It is also necessary to propagate changes to the access policy of such information. For example, EU data protection law includes ``secure delete'' and the ``right to be forgotten''.
This is achieved by capturing and recording the relationships between documents (and their system representations, \eg as files).
These complex relationships must be taken into account when deleting information or updating a policy.

\noindgras{Intrusion/Fault Detection:}
In cloud environments, several instances of the same system/application run in parallel. Through analysis of provenance data generated during those executions,  we can build a model of normal behavior.
By comparing the behavior of an executing task to this model, we can
detect abnormal behavior.
We discuss this use case in more detail in \autoref{sec:examples:intrusion}.

\noindgras{Fault Injection:}
Fault injection is a technique commonly used to test applications' resistance to failure~\cite{naughton2009fault}.
Colleagues at UCSC are using a CamFlow-generated provenance graph to record the 
execution of a distributed storage system to help them discover points of failure, leading to better fault-injection testing.

While CamFlow satisfies all of the above application requirements, in this paper, we focus on the construction of the capture mechanism itself.
In \autoref{sec:examples}, we discuss some of the provenance applications that we have built using CamFlow.

\section{Provenance Data Model} \label{sec:model}
We extend the W3C PROV Data Model (PROV-DM)~\cite{moreau2013prov}, which is used by most of the provenance community.
A provenance graph represents \emph{entities}, \emph{activities} and \emph{agents}. At the OS level, \emph{entities} are typically kernel data objects: files, messages, packets \etc, but also xattributes, inode attributes, exec parameters, network addresses \etc 
\emph{Activities} are typically processes carrying out manipulations on entities.
\emph{Agents} are persons or organisations (\eg users, groups \etc) on whose behalf the activities operate. 
In the provenance graph, all these elements are nodes, connected by edges representing different types of interactions.
For example, \autoref{image:introduction:w3c} depicts that an output file \emph{was generated by} a process that \emph{was associated with} a user Aline, and \emph{used} an executable and an input file.

CamFlow records how information is exchanged within a system through system calls.
Some calls represent an exchange of information at a point in time
\eg \texttt{read}, \texttt{write}; others may create shared state, \eg \texttt{mmap}.
The former can easily be expressed within the PROV-DM model, the latter is more complex to model.

Previous implementations managed shared states (\ie POSIX shared memory and mmap files, which both use the same underlying Linux kernel mechanism) by recording the action of ``associating'' the shared memory with a process (\eg \texttt{shmget}, \texttt{mmap\_file}).
These shared states are either ignored during queries or make the queries significantly more complex.
Indeed, information may flow implicitly (from the system perspective) between processes, without being represented in the provenance graph.
Without a view of information flow within an application (we discuss how intra-application provenance can be appended to the graph in \autoref{sec:application}), it is impossible to know if/when a process is writing to some ``associated'' shared memory at the system-call level.
Therefore, we conservatively assume that any incoming or outgoing flow to/from the process implies an incoming or outgoing flow to/from the shared state.

\begin{figure}[t]
\centering
  \includegraphics[width=\columnwidth]{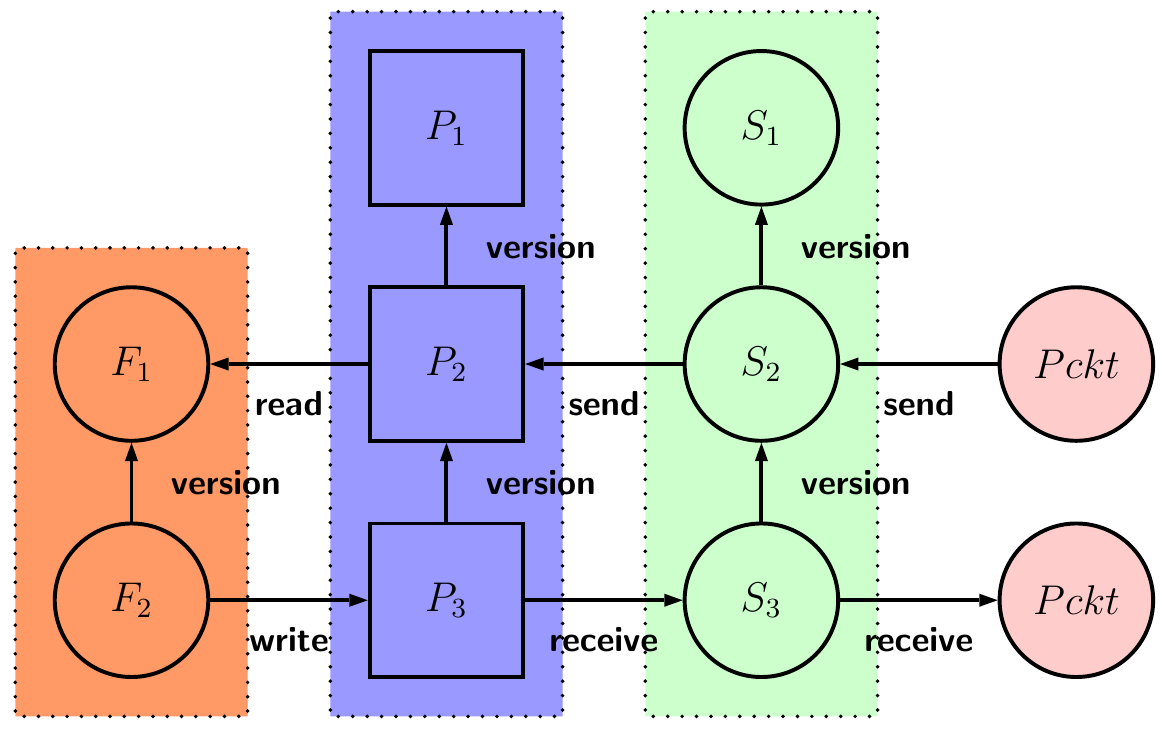}
  \caption{CamFlow-Provenance partial graph example. A process $P$ reads information from a file $F$, sends the information over a socket $S$, and updates $F$ based on information received through $S$.} 
  \label{image:kernel:graph}
\end{figure}

Provenance graphs are acyclic.
A central concept of a CamFlow graph is the \emph{state-version} of kernel objects, which guarantees that the graph remains acyclic.
A kernel object (\eg an inode, a process \etc) is represented as a succession of nodes, which represent an object changing state.
We conservatively assume that any incoming information flow generates a new object state.
Nodes associated with the same kernel object share the same object id, machine id, and boot id (network packets follow different rules described in \autoref{sec:cross}).
Other node attributes may change, depending on the incoming information (\eg \texttt{setattr} might modify an \texttt{inode's mode}).
\autoref{image:kernel:graph} illustrates CamFlow versioning.

In \autoref{image:kernel:graph}, we group
nodes belonging to the same \emph{entity} or \emph{activity} ($F$, $P$ and $S$),
based on shared attributes between those nodes (\ie \texttt{boot\_id}, \texttt{machine\_id}, and \texttt{node\_id}).
In the cloud context, those groups can be further nested into larger groups representing individual machines or particular clusters of machines.
For example, when trying to understand interactions between Docker containers~\cite{docker}, we can create groups based on namespaces (\ie UTS, IPC, mount, PID and network namespaces), control groups, and/or security contexts.
The combination of these process properties differentiate processes belonging to different Docker containers.
Provenance applications determine their own meaningful groupings.

\begin{lstlisting}[language=json, style=mystyle, caption={PROV-JSON formatted inode entry.}, label=listing:model:inode]
"ABAAAAAAACAe9wIAAAAAAE7aeaI+200UAAAAAAAAAAA=": {
    "cf:id": "194334",
    "prov:type": "fifo",
    "cf:boot_id": 2725894734,
    "cf:machine_id": 340646718,
    "cf:version": 0,
    "cf:date": "2017:01:03T16:43:30",
    "cf:jiffies": "4297436711",
    "cf:uid": 1000,
    "cf:gid": 1000,
    "cf:mode": "0x1180",
    "cf:secctx": "unconfined_u:unconfined_r:unconfined_t:s0-s0:c0.c1023",
    "cf:ino": 51964,
    "cf:uuid": "32b7218a-01a0-c7c9-17b1-666f200b8912",
    "prov:label": "[fifo] 0"
}

\end{lstlisting}

\begin{lstlisting}[language=json, style=mystyle, caption={PROV-JSON formatted write entry.}, label=listing:model:write]
"QAAAAAAAQIANAAAAAAAAAE7aeaI+200UAAAAAAAAAAA=": {
    "cf:id": "13",
    "prov:type": "write",
    "cf:boot_id": 2725894734,
    "cf:machine_id": 340646718,
    "cf:date": "2017:01:03T16:43:30",
    "cf:jiffies": "4297436711",
    "prov:label": "write",
    "cf:allowed": "true",
    "prov:activity": "AQAAAAAAAEAf9wIAAAAAAE7aeaI+200UAQAAAAAAAAA=",
    "prov:entity": "ABAAAAAAACAe9wIAAAAAAE7aeaI+200UAQAAAAAAAAA=",
    "cf:offset": "0"
 }
\end{lstlisting}

\noindent \autoref{listing:model:inode} and \autoref{listing:model:write} 
show an inode and a write relationship in the provenance graph, 
in PROV-JSON~\cite{Huyng2013provjson} format.\footnote{See \url{https://github.com/CamFlow/CamFlow.github.io/wiki/JSON-output}.}
We also developed a compact binary format.

\section{Capturing System Provenance} \label{sec:capture}
\begin{figure*}[!ht]
\centering
  \includegraphics[width=\textwidth]{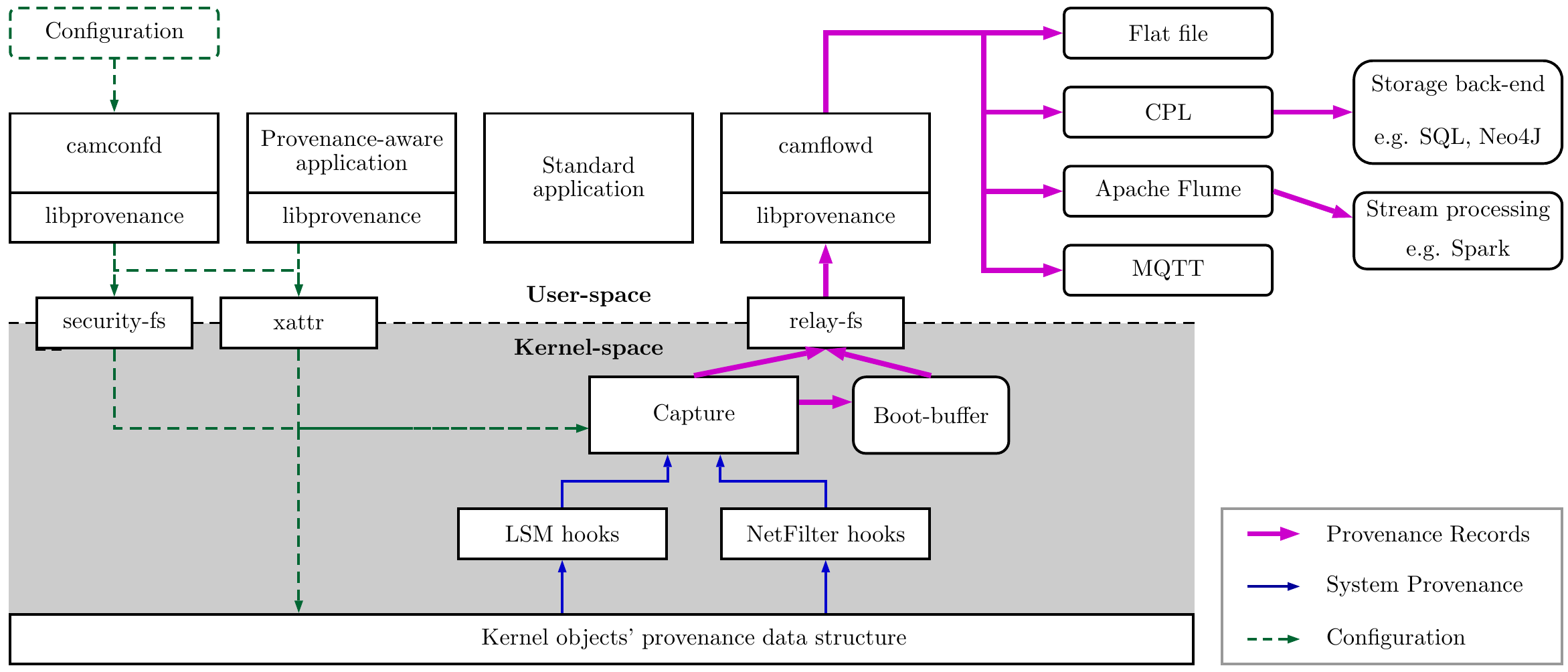}
  \caption{Architecture overview.} 
  \label{image:kernel:architecture}
\end{figure*}

CamFlow builds upon and learns from previous OS provenance capture mechanisms, namely PASS~\cite{muniswamy2006provenance, muniswamy2009layering}, Hi-Fi~\cite{pohly2012hi} and LPM~\cite{bates2015trustworthy}.
Our strategy for CamFlow is that it:
1) is easy to maintain,
2) uses existing kernel mechanisms whenever possible,
3) does not duplicate existing mechanisms, and
4) can be integrated into the mainline Linux kernel.
Beyond being good practice, the first three requirements are
essential to realize the last requirement (adoption in the mainline Linux kernel).
We also incorporated requirements for usability and extendibility.

\autoref{image:kernel:architecture} gives an overview of the architecture.
We use LSM and NetFilter (NF) hooks to capture provenance and
 \texttt{relayfs}~\cite{zanussi2003relayfs} to transfer the data to user space
where it can be stored or analyzed.
Applications can augment system level provenance with application-specific details through a pseudofile interface.
This is restricted to the owner of the capability \texttt{CAP\_AUDIT\_WRITE},
following Linux development recommendations to re-use existing capabilities,
rather than from creating new ones.
Applications owning the \texttt{CAP\_AUDIT\_CONTROL} capability can collect
configure CamFlow to capture a subset of
the provenance (see \autoref{sec:tailoring}).
We provide a library whose API
abstracts interactions with \texttt{relayfs} and the pseudo-files we use to
communicate configuration information.

\subsection{Capture mechanism}
\label{sec:kernel:capture}

The key to clean Linux integration is use of the LSM API.
LSM implements two types of security hooks that are called 1) when a kernel object (\eg \texttt{inodes}, \texttt{file}, \texttt{sockets} \etc) is allocated and 2) when a kernel object is accessed.
The first allocates a security blob, containing information about the associated kernel object.
The second is called when making access decisions during a system call (\eg \texttt{read}, \texttt{write}, \texttt{mmap\_file}).

There are two types of LSM: \emph{major} and \emph{minor} modules.
\emph{Major} modules use a security blob pointer (\eg \texttt{cred\nobreakdash->security}, \texttt{inode\nobreakdash->i\_security} \etc).
Examples include \texttt{SELinux}~\cite{smalley2001implementing} and \texttt{AppArmor}~\cite{bauer2006paranoid}.
\emph{Minor} security modules, such as \texttt{SMACK}~\cite{Corbet2007} and \texttt{LoadPin}~\cite{Corbet2016}, do not use such pointers.
Kernel version 4.1 allows a major module to be stacked with any number of minor modules.
After Linux performs its normal access control checks, it calls
LSM hooks in a pre-defined order.
If any LSM hook call fails, the system call immediately fails.
We arrange for the CamFlow hook to be called last
to avoid recording flows that are subsequently blocked by another LSM.
CamFlow users the record of LSM events to build its
provenance graph~\cite{pohly2012hi}.

CamFlow needs a ``provenance'' blob pointer to store provenance information alongside kernel objects, so CamFlow is a major security module.
Using CamFlow in conjunction with another major module, e.g.,
SELinux, requires that we modify some kernel data structures, e.g.,
\texttt{inode}, \texttt{cred}, to add an additional pointer.
This is the one place we violate our ``self-contained'' philosophy. 
However, work is in progress~\cite{Schaufler2016} to allow stacking of
\emph{major} security modules (by modifying how security blobs are handled).

CamFlow hooks: 1) allocate a provenance-blob, if needed 2) consider filter/target constraints (see \autoref{sec:tailoring}), and 3) record provenance in the \texttt{relayfs} buffer.
The user space provenance service then reads the
provenance entries see \autoref{sec:kernel:interface}).

\subsection{Cross-host Provenance Capture}
\label{sec:cross}

CamFlow captures incoming packets through the \texttt{socket\_sock\_rcv\_skb} LSM hook.
No equivalent hook exists for outgoing packets, so we implement a \texttt{NetFilter} hook to capture their provenance.

CamFlow represents a packet as an entity in the provenance graph.
We construct packet identifiers from the immutable elements of the IP packet and immutable elements of the protocol (\eg TCP or UDP).
For example, a TCP packet's identifier is built from: IP packet ID, sender IP, receiver IP, sender port, receiver port, protocol, and TCP sequence number.
We also record additional metadata such as the payload length.
In most scenarios, the sending and receiving machines can match packets using
this information.
In other scenarios (\eg NAT), some post-processing may be required to match packets (\eg content match, temporal relationship, approximative packet length); the details of these techniques are beyond the scope of this paper.
Additionally, CamFlow can optionally capture payloads.

\begin{figure}[t]
\centering
  \includegraphics[width=\columnwidth]{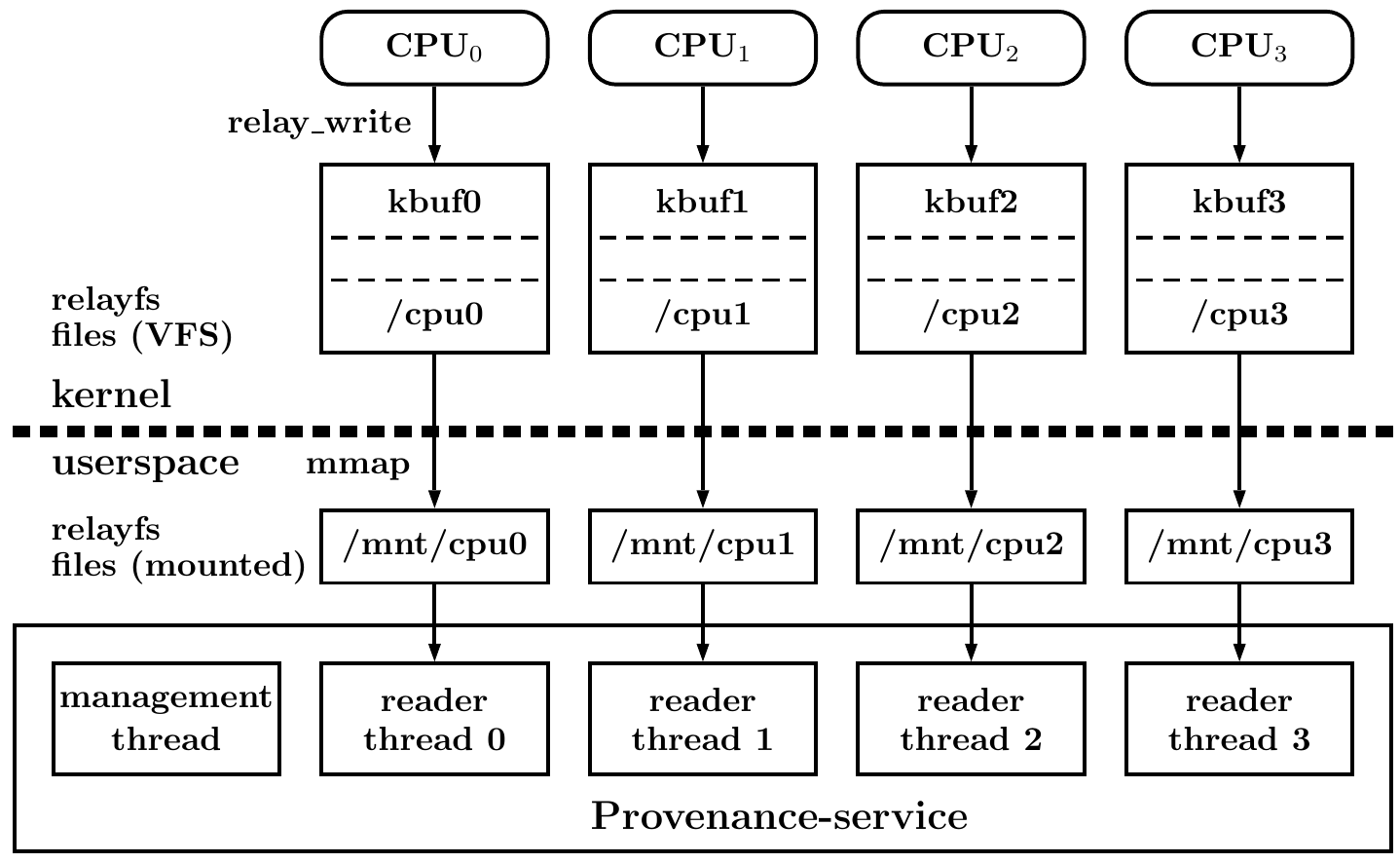}
  \caption{Transferring provenance data to user space through \texttt{relayfs}.}
  \label{image:kernel:relay}
\end{figure}

\subsection{Hardware Root of Trust}
\label{sec:kernel:trust}

In some circumstances it is necessary to be able to vouch for the integrity of provenance data and the provenance capture mechanism, \eg when presented as evidence in a court of law. The software may not always be under the control of a trusted entity, and hardware-based trust certification will be required.
Our architecture can support this by using 
Trusted Platform Module (TPM)~\cite{morris2011trusted} or virtual TPM~\cite{perez2006vtpm} technologies that can be used to measure the integrity of a system.
This integrity can be verified through Remote Attestation~\cite{kil2009remote}.

We need to assume that sufficient technical and non-technical measures are taken to ensure that the hardware has not been tampered with (\eg see CISCO vs NSA~\cite{rubinstein2014privacy}). 
We assume that the integrity of the low-level software stack is recorded and monitored at boot time through a Core Root of Trust Measurement (CRTM). 
The low-level software stack here includes: BIOS, boot loader code and configuration, options ROM, host platform configuration, \etc\@
We assume that this integrity measurement is kept safe and cannot be tampered with (one of the features provided by TPM).
Further, system integrity needs to be guaranteed at run time.

In the Linux world such the Integrity Measurement Architecture (IMA)~\cite{sailer2004design, jaeger2006prima, kil2009remote}
provies integrity guarantees and remote attestation.
The CRTM measures hardware and kernel integrity.
IMA is initialized before kernel modules are loaded and before any application is executed.
IMA measures the integrity of kernel modules, applications executed by root, or libraries.
Each binary is hashed to create an evidence trail. 
The integrity of this record is guaranteed by a cumulative hash, which is stored in TPM's Platform Configuration Registers.
During remote attestation, this evidence trail is verified against the known good value.
While not strictly part of CamFlow, use of IMA and remote attestation should be considered by cloud providers. 

\subsection{User Space Interface}
\label{sec:kernel:interface}

While provenance capture must happen in the kernel, we relegate storage and
processing to user space.
CamFlow uses \texttt{relayfs}~\cite{zanussi2003relayfs} to efficiently transfer data from kernel to user space as illustrated in \autoref{image:kernel:relay}.
The CamFlow kernel module writes 
provenance records into a ring buffer that is mapped to a pseudofile (this is handled by \texttt{relayfs})
that can be \texttt{read} or \texttt{mmapped} from user space.
Before file system initialization, CamFlow records provenance into a ``boot'' buffer that is copied to \texttt{relayfs} as soon as it is initialized.

Once the file system is ready, \texttt{systemd} starts up a provenance service in user space.
The service reads from the file that is mapped to the kernel buffer, and the \texttt{relayfs} framework deals with ``freeing'' space in the kernel ring buffer.
The provenance service itself is marked as \emph{opaque}
to avoid the provenance service looping infinitely;
the provenance capture mechanism ignores opaque entries, which do not appear in the captured data (see \autoref{sec:tailoring}).

We provide a user space library that deals with handling the \texttt{relayfs} files and multi-threading.
The library also provides functions to serialize the efficient kernel binary encoding into PROV-JSON.
We then stream the provenance data, in either binary or JSON format,
to applications, as shown in \autoref{image:introduction:prov}.
CamFlow can use any messaging middleware such as MQTT~\cite{banks2014mqtt} or Apache Flume~\cite{hoffman2013apache} to stream the provenance data.

\subsection{Application-level Provenance}
\label{sec:application}

So far we have discussed how CamFlow records information flows through system calls, in terms of kernel objects.
Other systems (e.g., database systems, workflow systems) record provenance in terms of different objects, e.g., tuples in a database system~\cite{buneman2007provenance} or stages of processing in a workflow system~\cite{crawl2011provenance, akoush2013hadoopprov}.
Muniswamy-Reddy \etal~\cite{muniswamy2009layering} demonstrated that many scenarios require associating objects from these different layers of abstraction to satisfy application requirements.

Consider scientific software reading several input files to generate several figures as output files.
Application-level provenance is oblivious to files altered by other applications and can only provide an incomplete view of the system.
On the other hand, system-level provenance, while providing a complete view of the system, will expose only coarse grained dependencies (\ie every output will depend on all inputs read so far).
Combining application- and system-level provenance allows us to describe fine-grained dependencies between inputs and outputs.

CamFlow provides an API for provenance-aware applications to disclose provenance
to support the integration of application and system provenance.
The API allows us to associate application provenance with system objects as long as a system level descriptor (\eg a file descriptor) to that object is available to the application. 
CamFlow guarantees that such associations do not produce cycles, however, it is the responsibility of the application level provenance collection to ensure that there are no cycles within its subgraph.
Further, CamFlow's underlying infrastructure manages identifiers and guarantees their uniqueness, system wide.

Application-level provenance associated with a kernel object is attached to a specific version of that object in the provenance graph.
It is sometimes useful to view
application-disclosed provenance as a subgraph contained within \emph{activity} nodes, describing the internal working of that particular \emph{activity}.

\begin{figure}[t]
	\centering
	\includegraphics[width=\columnwidth]{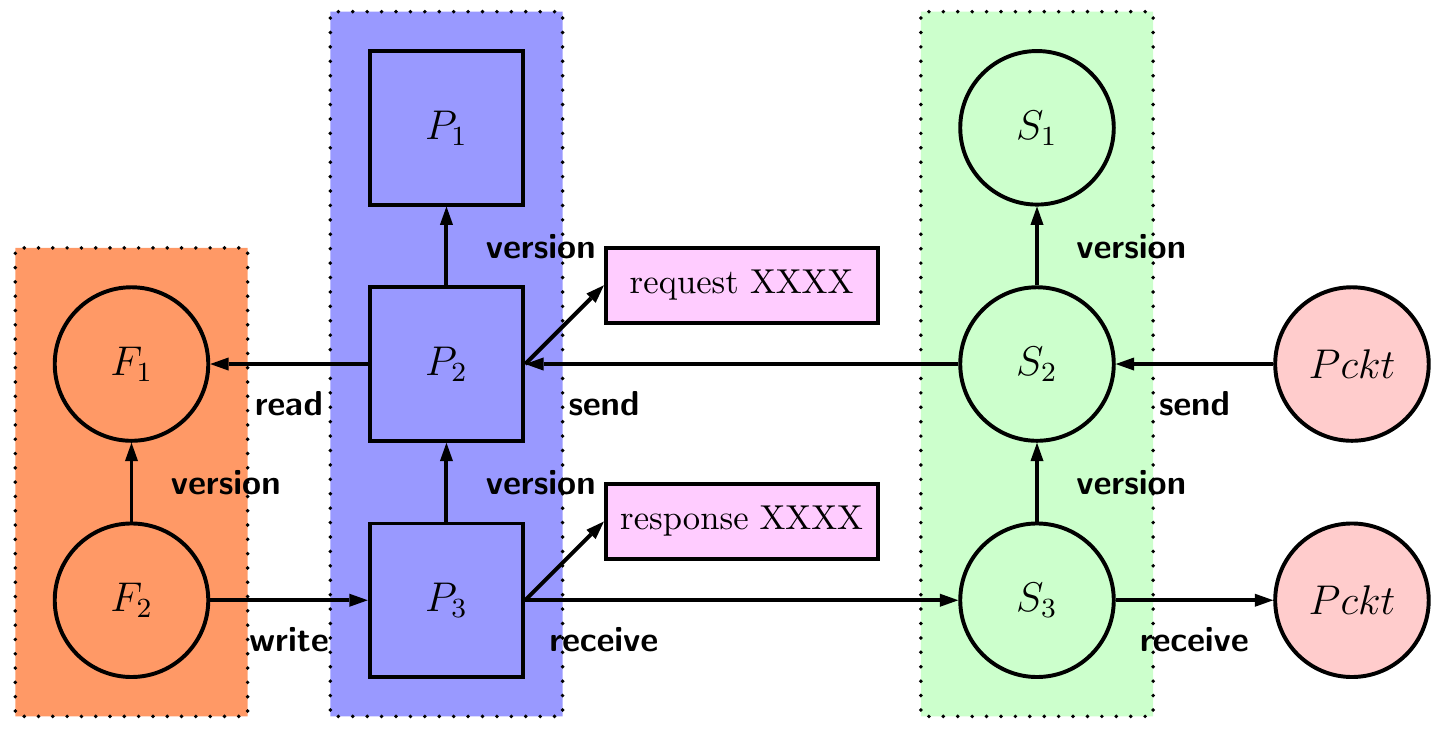}
	\caption{CamFlow-Provenance presented in \autoref{image:kernel:graph} annotated with application logs.} 
	\label{image:kernel:annotated}
\end{figure}

Modifying existing applications to be provenance-aware is not a trivial task~\cite{muniswamy2009layering, lerner2014rdatatracker}.
Ghoshal \etal~\cite{ghoshal2013provenance} demonstrated that existing application log files can be transformed into a provenance graph.
We provide the infrastructure for application to apply such approach.

We limit this approach to applications that already generate log files (\eg \texttt{httpd}, \texttt{Postgres}).
Placing these logs in a special pseudo-file automatically incorporates their log records into CamFlow's provenance graphs.
For example, replacing \texttt{ErrorLog "/var/log/httpd-error.log"} with \texttt{ErrorLog "/sys/kernel/security/provenance/log"} in the \texttt{httpd} configuration file enables CamFlow to incorporate web server operations.
CamFlow creates an entity for each log record and attaches it to the appropriate version of activity that created it.
The log record entities contain the text of the log message, thereby providing annotations for the provenance graph.
This helps make the graph more human-readable.
If desired, provenance applications can parse these log entries for further processing or interpretation.

\autoref{image:kernel:annotated} shows a log integrated into the provenance graph. In this example, the log entries might help a human user or application understand that the information exchanged with the socket corresponds to a client request/response interaction.
In \autoref{sec:examples:server}, we use this feature to insert httpd logs into a system provenance graph.

\section{Tailoring Provenance Capture} \label{sec:tailoring}
\begin{lstlisting}[language=Ini, style=mystyle, caption={Capture policy example.}, label=listing:kernel:capturepolicy, float,floatplacement=t]
[provenance]
;unique identifier for the machine, use hostid if set to 0
machine_id=0
;enable provenance capture
enabled=true
;record provenance of all kernel object
all=false
;set opaque file
opaque=/usr/bin/ps
;set tracked file
;track=/home/thomas/test.o
;propagate=/home/thomas/test.o
node_filter=directory
node_filter=inode_unknown
node_filter=char
relation_filter=sh_read
relation_filter=sh_write
propagate_node_filter=directory
propagate_node_filter=char
propagate_node_filter=inode_unknown

[ipv4-egress]
;propagate=0.0.0.0/0:80
;propagate=0.0.0.0/0:404
;record exchanged with local server
;record=127.0.0.1/32:80

[user]
;track=thomas

[group]
;propagate=docker
\end{lstlisting}

A major concern about whole-system provenance capture is that the size of the
provenance can ultimately dwarf the ``primary'' data.
Further, provenance application execution time is often a function of the size of the provenance graph.
One approach is to limit provenance capture to ``objects of interest'', based on a particular security policy~\cite{bates2015take,pasquier2016ic2e,pasquier2015camflow}.
While recording interactions only between security-sensitive objects may make sense in some contexts, in others, for example, a scientific workflow, 
it probably does not make sense.
We devised a more complex capture policy that allows end users to define the scope of capture based on their precise needs, greatly reducing storage overhead.
Users choose between whole-system provenance and selective provenance.
Within selective provenance, they can specify filters on nodes and edges, specific programs or directories to capture, whether capture should propagate from a specified process, and what network activity should be captured.
Users specify these capture policies either through a command line interface or
in a policy configuration file, like the one shown in 
\autoref{listing:kernel:capturepolicy}.

In whole-system provenance mode, CamFlow captures all objects that
have not been declared opaque (line 7).
In selective mode (line 5), CamFlow captures the provenance of only non-opaque specified targets; line 9,
\texttt{track=/home/thomas/myexperiment.o}, tells CamFlow to track any information flow to/from the specified file and any process resulting from the execution of programs built from it.
Line 10 tells CamFlow to track any object that interacts with the file/process and any object they may subsequently interact with.
Similarly, line 27, \texttt{propagate=0.0.0.0/0:0}, indicates that we want to track any information flow to/from a listening socket on the machine.

\begin{figure}[t]
	\centering
	\includegraphics[width=\columnwidth]{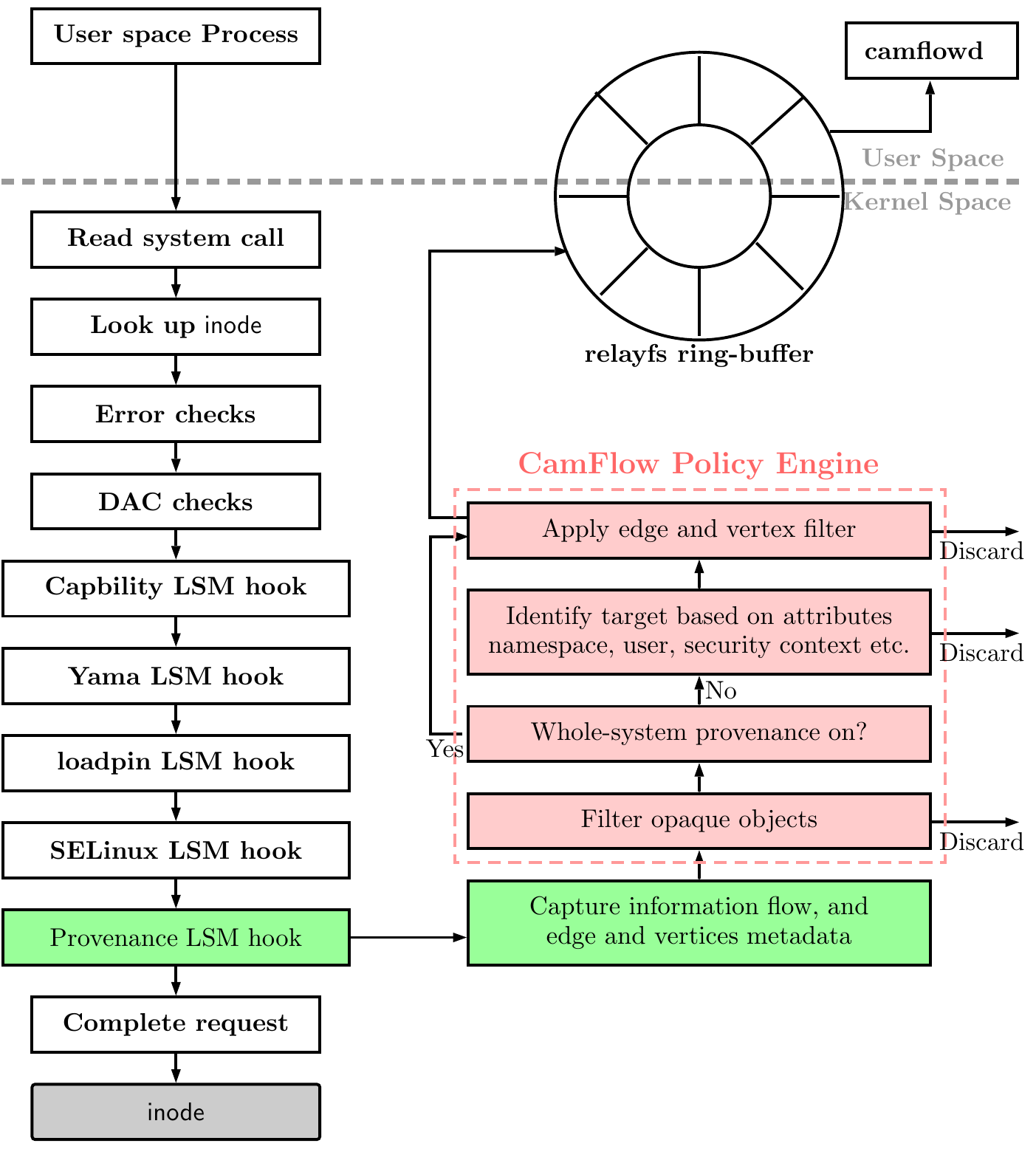}
	\caption{Executing an \texttt{open} system call. In green the capture mechanism described in \autoref{sec:capture}, in pink the provenance tailoring mechanism described in \autoref{sec:tailoring}.} 
	\label{image:tailoring}
\end{figure}

We can also specify opaque or tracked objects based on criteria, such as:
1) pathname (line 9), 2) network address (line 27), 3) LSM security context\footnote{These are dependent on the major LSM running on a machine. The security contexts defined by SELinux or AppArmor are different. Defining capture policy based on these, therefore requires an understanding of the MAC mechanism in place.}
(not shown),
4) control group,\footnote{Control groups or \texttt{cgroup}s are the mechanism used in Linux to control, for example, the resource access of containers.}
(not shown), and
5) user and group ID.
We frequently expand this list based on feedback from users.
\autoref{image:tailoring} illustrates how CamFlow implements the capture policy for the \texttt{open} system call.

The mechanism to \emph{mark} targets varies dependending on the targeted attribute. For example, exact path matches are handled through security extended attributes, while network interface based policies are handled in \texttt{connect} and \texttt{bind} LSM hooks.
The policy engine does not reevaluate the policy for every event, but rather verifies the mark associated with kernel objects and the provenance data structure.
Some policies (\eg user id) are regularly reevaluated, for example, on \texttt{setuid} events.

\noindgras{Excluding edges or node types from the graph:}
Line 12 (14) illustrates how to exclude specific node (edge) types.
In this case, line 12 tells CamFlow not to capture operations on directories,
while line 14 says to ignore read permission checks.
This is the last stage in an event capture, as illustrated in \autoref{image:tailoring}.
For every edge recorded, CamFlow compares the types of its endpoints to that specified in the filter, discarding any edge that produced a match.

\noindgras{Propagating tracking: }
While targeting a single individual or group of objects may be interesting, in many cases the user wants to track all information emanating from a given source.
The \texttt{propagate} setting tells CamFlow to track interactions with any object that received information from a tracked object; this automatically tracks the receiving object in a manner similar taint tracking~\cite{enck2014taintdroid} works.
It is also possible to define types of relations or nodes through which tracking should not propagate (\eg \texttt{propagate\_node\_filter=directory}).
Filtered nodes or relation types do not propagate tracking.
For example, when tracking content dissemination (see \autoref{sec:examples:DLP}) it is possible to set the tracking propagation to never be applied to \texttt{directories}, or not to occur on permissions-checking operations (\eg when opening a file), but only on explicit \texttt{read} and \texttt{write} operations.

\noindgras{Taint tracking: }
Tracking propagation can be coupled with tainting (with a taint being
an arbitrary 64 bit integer), following the same policy-specified restrictions.
Kernel objects can hold an arbitrary number of taints.
This is useful when provenance is used for policy enforcement, requiring a quick decision to be made upon accessing an object (see \autoref{sec:examples:DLP}).

Note that ``tailoring'' provenance capture means that
the provenance is no longer guaranteed to be \emph{complete}.
Indeed, the \emph{raison d'\^etre} for tailoring is to exclude part of the provenance from the record.
Thus, it is imperative that a user understand the \emph{minimum} and \emph{sufficient} information required for
a particular application to be \emph{correct}.

We can, for example, make the distinction 
between \emph{data provenance}~\cite{carata2014primer} and \emph{information flow audit}~\cite{pasquier2016ic2e}.
The two are similar, but differ in their objectives.
The first is concerned with where data comes from; 
that is, the \emph{integrity} of the data (\eg verifying properties of data used for a scientific results).
The second is concerned with where the data have gone; 
that is, the \emph{confidentiality} of the data (\eg monitoring usage of personal data in a system).
In terms of \emph{completeness}, assuming an object or objects of interest,
this means that for data provenance (integrity), all information flows \textbf{to} the object 
must be recorded; and for infomation flow audit (confidentiality), all information flows \textbf{from} the object  must be recorded.
The tracking propagation mechanism, for example, is appropriate to handle \emph{confidentiality}, but not necessarily \emph{integrity}.
It is important to define the \emph{objective} of the capture and the \emph{adversary} (if any) when tailoring the capture.

\section{Evaluation} \label{sec:evaluation}
The goal of our evaluation is to
answer the following questions:
1) How does CamFlow improve maintainability relative to prior work?
2) What is the impact of CamFlow on PaaS application performance?
3) How much does selective provenance capture reduce the amount of data generated?
We follow this quantitative evaluation with more qualitative evaluation of
usability in \autoref{sec:examples}.

\subsection{Maintainability}
\label{sec:eval:maintainability}

\begin{table}[h]
\resizebox{\columnwidth}{!}{
\begin{tabular}{ l | c c c | c}
  Provenance system & Headers & Codes & Total & LoC \\
  \hline
  (kernel version) & \multicolumn{3}{c}{Number of files} & \\
  \hline
  PASS (v2.6.27.46) & 18 & 69 & 87 & 5100\\
  LPM (v2.6.32 RHEL 6.4.357)& 13 & 61 & 74 & 2294\\
  CamFlow (v4.9.5) & 8 & 1 & 9 & 2428\\
\end{tabular}
}
\caption{Number of files and lines of code (LoC) modified in various
whole-system provenance solutions.
The LoC results for PASS and LPM are as reported in~\cite{muniswamy2006provenance} and~\cite{bates2015trustworthy} respectively. The number of files modified measurement is based on the last source code available from the PASS and Hi-Fi/LPM teams and for CamFlow v0.2.1.}
\label{table:eval:maintain}
\end{table}

It is difficult to objectively measure the maintainability of a code base.
PASS was one of the first whole-system provenance capture mechanisms, but it required massive changes to the Linux kernel, and it was unmaintainable, almost from day one.
We never simply upgraded PASS across Linux kernel minor versions, let alone major versions, because it was such a massive undertaking.
It took us over two years and several dozen person months to develop the second version of PASS~\cite{muniswamy2009layering}, which included a transition from Linux 2.4.29 to 2.6.23, i.e., two minor Linux revisions.
Our estimate is that the effort was divided in roughly three equal parts: re-architecting how we interacted with the kernel, adding new features, and changing kernel releases.
It should be apparent that maintaining a self-contained provenance capture module will be significantly easier than maintaining PASS, and the ease with which we have adapted to new Linux releases supports this.
Indeed, an explicit requirement of the LSM framework~\cite{morris2002linux} is to limit the impact of new security features on the kernel architecture, thereby improving the maintanability of the Linux kernel.
CamFlow selected the LSM framework for precisely this reason.
Further as LSM is a major security feature of the Linux kernel to our knowledge most Linux distribution ship with an LSM (\eg SELinux for Android >=4.3),
it is practically guaranteed that future releases will continue to support
the framework with minimal modification.

\autoref{table:eval:maintain} provides some metrics to quantify the pervasiveness required for PASS, LPM, and CamFlow.
We calculated the numbers using
\texttt{diff} between ``vanilla'' Linux and provenance-enabled Linux for the same version, ignoring newly added files.

\begin{table}[h]
\resizebox{\columnwidth}{!}{
\begin{tabular}{ l | c c c | c c}
  Kernel version & Code & Header & CamFlow & Total & Conflict \\
  \hline
  & \multicolumn{3}{c}{In number of files modified} & \multicolumn{2}{c}{In number of lines modified}\\
  \hline
  4.9.6 $\rightarrow$ 4.10-rc6 & 0 & 1 & 0 & 234 & 0 \\
  4.9.5 $\rightarrow$ 4.9.6 & 0 & 0 & 0 & 0 & 0 \\
  4.9.4 $\rightarrow$ 4.9.5 & 0 & 0 & 0 & 0 & 0 \\
  4.9.2 $\rightarrow$ 4.9.4 & 0 & 0 & 0 & 0 & 0 \\
  4.9 $\rightarrow$ 4.9.2 & 0 & 0 & 0 & 0 & 0 \\
  4.4.38 $\rightarrow$ 4.9 & 1 & 4 & 3 & 1194 (17) & 8 \\
  4.4.36 $\rightarrow$ 4.4.38 & 0 & 0 & 0 & 0 & 0 \\
  4.4.35 $\rightarrow$ 4.4.36 & 0 & 0 & 0 & 0 & 0 \\
  4.4.34 $\rightarrow$ 4.4.35 & 0 & 0 & 1 & 2 (2) & 0 \\
  4.4.32 $\rightarrow$ 4.4.34 & 0 & 0 & 0 & 0 & 0 \\
\end{tabular}
}
\caption{Modification required when porting CamFlow to new kernel versions. The number of lines modified includes comments. The numbers in parentheses are the number of lines of CamFlow-specific code changed.}
\label{table:eval:update}
\end{table}

Neither PASS nor LPM were ever ported to newer kernel releases, even though both groups would have liked to do so.
In contrast, we have already maintained CamFlow across multiple Linux kernel major revisions with only a few hours of effort.
The authors of PASS and LPM state that transitioning to newer kernel releases would take many man months.
\autoref{table:eval:update} quantifies the CamFlow porting effort.
We obtained these results by going through the updates in our git repository from November 21, 2016 to February 2, 2017.
We considered only changes relating to version updates and ignored feature development.
The most significant upgrade
over this period was from version 4.4.38 to 4.9.
However, we modified 8 lines of code in \texttt{/security/security.c} relating to changes in the LSM framework.
Most of these patches were applied without any conflict.
We continue to maintain CamFlow, see \autoref{sec:availability} for the most recent supported version and a full commit history. (Neither PASS nor LPM have any such commit history available.)

\subsection{Performance}
\label{sec:eval:performance}

We next discuss how CamFlow affects system performance.
We selected standard benchmarks that can be easily reproduced and provide meaningful references to prior work.
We ran the benchmarks on a bare metal machine with 6GiB of RAM and an Intel i7-4510U CPU.
(We use a bare metal machine rather than a cloud platform, such as EC2, to reduce uncontrolled variables.)
See \autoref{sec:availability} for details on obtaining the code, reproducing these results and accessing the raw results presented here,
following suggestions from Collberg et. al~\cite{collberg2016repeatability}.

We run each test on three different kernel configurations.
The \emph{vanilla} kernel configuration is our baseline and corresponds to an unmodified Linux kernel. 
The \emph{whole} configuration corresponds to CamFlow recording whole-system provenance.
The \emph{selective} configuration corresponds to CamFlow with whole-system provenance off, but selective capture on.
We use
CamFlow v0.2.1~\cite{thomas_pasquier_2017_571427} with Linux 4.9.5.
Comparing the CamFlow configurations to the baseline reveals the overhead imposed by CamFlow's provenance capture mechanism.

\begin{table}[h]
\resizebox{\columnwidth}{!}{
\begin{tabular}{ l | c c c c c}
  Test Type & vanilla & whole & overhead & selective & overhead \\
  \hline
  \multicolumn{6}{c}{Process tests, times in $\mu s$, smaller is better}\\
  \hline
  NULL call			& 0.07 	& 0.06 	& 0\% 	& 0.05 	& 0\%\\
  NULL I/O			& 0.09 	& 0.14 	& 56\% 	& 0.14 	& 56\%\\
  stat 				& 0.39 	& 0.78 	& 100\% & 0.50 	& 28\%\\
  open/close file 	& 0.88 	& 1.59 	& 80\% 	& 1.04 	& 18\%\\
  signal install 	& 0.10 	& 0.10 	& 0\% 	& 0.10 	& 0\%\\
  signal handle 	& 0.66 	& 0.67 	& 2\% 	& 0.66 	& 0\%\\
  fork process		& 110 	& 116 	& 6\% 	& 112	& 2\%\\
  exec process 		& 287 	& 296 	& 3\% 	& 289 	& <1\%\\
  shell process 	& 730 	& 771 	& 6\% 	& 740 	& 1\%\\
  \hline
  \multicolumn{6}{c}{File and memory latencies in $\mu s$, smaller is better}\\
  \hline
  file create (0k)	& 5.05 	& 5.32	& 5\%	& 5.12 & 1\%\\
  file delete (0k)	& 3.42 	& 3.78 	& 11\% 	& 3.79 & 11\%\\
  file create (10k)	& 9.81 	& 11.41	& 16\% 	& 10.9 & 11\%\\
  file delete (10k)	& 5.70 	& 6.12 	& 7\% 	& 6.08 & 6\%\\
\end{tabular}
}
\caption{LMbench measurements.}
\label{table:eval:lmbench}
\end{table}

\begin{figure}[t]
\centering
  \includegraphics[width=\columnwidth]{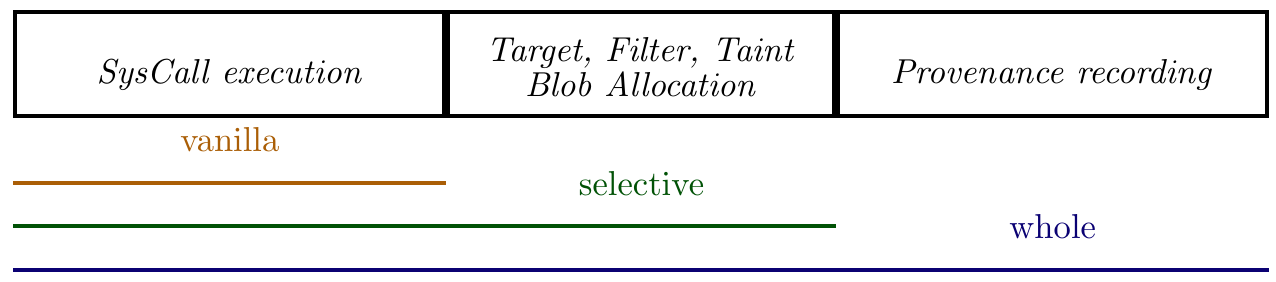}
  \caption{System call overhead decomposition.} 
  \label{image:eval:decomposition}
\end{figure}

\noindgras{Microbenchmarks: }
We used LMbench~\cite{mcvoy1996lmbench} to benchmark the impact of CamFlow's provenance capture on system call performance.
\autoref{table:eval:lmbench} presents a subset of the results.
Due to space constraints, we selected the subset of greatest relevance;
the complete results are available online.

The performance overhead can be decomposed into two parts as illustrated in \autoref{image:eval:decomposition}.
The first part is present in the \emph{selective} and \emph{whole} results.
This part concerns the allocation of the provenance blob associated with kernel objects (see \autoref{sec:capture}) and the application of the policies described in \autoref{sec:tailoring}.
The second part is present in the \emph{whole} overhead only and corresponds to the recording of the provenance.

It is important to note that a system call does not necessarily correspond to a single event in the provenance graph.
On a file open, for example, both \texttt{read\_perm} and \texttt{write\_perm} provenance events may be generated as appropriate for the directory and subdirectories in the path and for the file itself, in addition to an event for the \texttt{open} on the file.
Further, if the file did not exist prior to the \texttt{open}, CamFlow generates a \texttt{create} event.
Tailoring the types of provenance events that must be recorded (see \autoref{sec:tailoring}) can further improve performance.
Additionally, assuming a relatively constant time for provenance data allocation and copy into the relay buffer, the overhead appears proportionally large due to the short execution time of most system calls.
For example, \texttt{stat} is fast, but must record \texttt{perm\_read}, \texttt{read} and \texttt{get\_attr} provenance relations, producing large
\emph{relative} overhead, which in \emph{absolute} terms is smaller than that for \texttt{exec} (0.39$\mu$s vs 9$\mu$s for whole provenance), which requires several provenance object allocations and a larger number of relations.

\begin{table}[h]
\resizebox{\columnwidth}{!}{
\begin{tabular}{ l | c c c c c}
  Test Type & vanilla & whole & overhead & selective & overhead \\
  \hline
  \multicolumn{6}{c}{Execution time in seconds, smaller is better}\\
  \hline
  unpack 				& 11.16	& 11.36 & 2\% & 11.24 & <1\% \\
  build 				& 433	& 442 	& 2\% & 429 & 0\% \\
  \hline
  \multicolumn{6}{c}{4kB to 1MB file, 10 subdirectories,}\\
  \multicolumn{6}{c}{4k5 simultaneous transactions, 1M5 transactions}\\
  \hline
  postmark 				& 71 	& 79 	& 11\% & 75s & 6\%\\
  \hline
\end{tabular}
}
\caption{Standard provenance-system macrobenchmark results.}
\label{table:eval:macrobenchmark}
\end{table}

\noindgras{Macrobenchmarks: }
We present two sets of macrobenchmarks that
provide context for the overheads measured in the microbenchmarks.
First, we used the Phoronix test suite~\cite{larabel2011phoronix} to illustrate single machine performance, running kernel \emph{unpack} and \emph{build} operations.
Second, we ran the Postmark benchmark~\cite{katcher1997postmark}, simulating the operation of an email server.
These are the common benchmarks used in the system provenance literature.
Next, we ran a collection of benchmarks frequently used in cloud
environments, also available via the Phoronix test suite.
\autoref{table:eval:macrobenchmark} shows the results of the single system
benchmarks, and 
\autoref{table:eval:cloudybench} shows the results for the cloud-based
benchmarks.

\begin{table}[h]
\begin{center}
\resizebox{0.9\columnwidth}{!}{
\begin{tabular}{ l | c c c}
  Test Type & off & whole & overhead \\
  \hline
  \multicolumn{4}{c}{Request/Operation per second (the higher the better)}\\
  \hline
  apache 				& 8235	& 7269 & 12\% \\
  redis (LPOP)		    & 1442627	& 1120935 	& 22\%  \\
  redis (SADD)		    & 1144405	& 1068877 	& 7\%  \\
  redis (LPUSH)		    & 998077	& 961849 	& 4\%  \\
  redis (GET)		    & 1386329	& 1219830 	& 12\%  \\
  redis (SET)		    & 1053978	& 1012158 	& 4\%  \\
  memcache (ADD)		& 18297	& 16658 	& 9\%  \\
  memcache (GET)		& 37826	& 32548 	& 14\%  \\
  memcache (SET)		& 17762	& 16825 	& 5\%  \\
  memcache (APPEND)		& 19393	& 17172 	& 11\%  \\
  memcache (DELETE)		& 38245	& 32517 	& 15\%  \\
  php					& 240889 & 239069	& <1\% \\
  \hline
  \multicolumn{4}{c}{Execution time (ms) (the lower the better)}\\
  \hline
  pybench 				& 3515 	& 3525 	& <1\% \\
  \hline
\end{tabular}
}
\end{center}
\caption{Extended macrobenchmark results.}
\label{table:eval:cloudybench}
\end{table}

The single system benchmarks (\autoref{table:eval:macrobenchmark}) show that
CamFlow adds minimal overhead for the kernel unpack and build benchmarks
and as much as 11\% overhead to Postmark for whole-system capture.
These results compare favorably to prior systems and the option to use
selective capture provides a better performing solution.

We ran two configurations for the cloud-based benchmark: \emph{off} includes CamFlow in the kernel build but does not capture any data, representing tenants who do not use the provenance capture feature, and \emph{whole} runs in whole-system capture mode.
This time, the overheads are larger, with a maximum of 22\% for redis LPOP.

\noindgras{Discussion: }
CamFlow's performance is the same order of magnitude as that reported by PASS~\cite{muniswamy2006provenance}, Hi-Fi~\cite{pohly2012hi}, its successor LPM~\cite{bates2015trustworthy}, and SPADE~\cite{gehani2012spade}, and slightly better in most cases.
However, these results are difficult to compare---the variation in experimental setup is significant
\eg from a 500MHz Intel Pentium 3 with 768 MiB RAM for PASS to a dual Intel Xeon CPU machine for LPM,
or \eg kernel version from 2.4.29 for PASS to 4.9.5 for CamFlow.
Further the provenance ``coverage'' tends to become more extensive in each successive publication.
Comparing to the latest published work---LPM---CamFlow includes additional information such as \texttt{iattr} and \texttt{xattr}, \texttt{security context}, \texttt{control group}, \texttt{overlayfs} \etc
While this provides a more complete picture, it also increases the amount of provenance recorded, which produces more overhead.

\subsection{Generated Data Volume}
\label{sec:eval:storage}

CamFlow is a capture mechanism, so we do not attempt to evaluate provenance storage strategy, but rather the impact of selective capture on the amount of data generated.
As prior systems~\cite{muniswamy2006provenance, pohly2012hi, gehani2012spade, bates2015trustworthy} have shown, storage and the accompanying query time increase over time.
We discuss an approach to address increasing query time for some scenarios in \autoref{sec:examples:DLP}.

We currently write provenance in a bloated, but standard format (PROV-JSON).
Colleagues working with CamFlow have achieved a storage size of 8\% of the equivalent PROV-JSON graph size through compression techniques~\cite{provcompress}.
Others have reported on alternative ways to reduce storage requirements through compression ~\cite{chapman2008efficient,xie2011compressing,xie2012hybrid}, but such concerns are orthogonal to the work presented here. Regardless of the storage techniques employed, the size of the graph grows over time, and this makes query time grow proportionally.

The CamFlow capture mechanism addresses storage issues in the following ways:
1) it gives the storage system great flexibility through a clear separation of concerns;
and 2) it allows for configurable capture, limiting the volume of data.
The separation of concerns
is application specific and out of scope, however, previous work by Moyer \etal~\cite{Moyer16} discusses handling provenance storage in a cloud environment.
In the rest of this section we evaluate the impact of selective capture.

We selected a simple workload to ease reproducibility; we
record the provenance generated by:\\
    \texttt{wget www.google.com}\\
We have three experimental setups: 1) provenance off (baseline); 2) ignore directory nodes and permission edges (\texttt{perm\_read}, \texttt{perm\_write} and \texttt{perm\_exec}) (selective);
and 3) record everything (whole). Packet payloads are \emph{not} recorded in any of these setups.
Note that there are 39 edge types and 24 node types in the current implementation.
The data recorded when the provenance is off corresponds to the machine/boot description (<1kB).
The output rate is 12.45kB/iteration for \emph{whole} and 9.98kB/iteration for \emph{selective}; tailoring capture reduces the data rate by 20\%.
This illustrates the trade-off between between completeness and utility (see discussion in \autoref{sec:tailoring}).
In scenarios where, for example, only a set of applications or the actions of a user are of interest, the quantity of data captured can be several orders of magnitude smaller relative to PASS or LPM. As discussed in \autoref{sec:tailoring}, it is possible express in detail what part of the system is/is not of interest.
This trade-off needs to be viewed in relation to the provenance application;
we discuss a concrete example in \autoref{sec:examples:DLP}.

\subsection{Completeness and expressiveness}
\label{sec:eval:discussion}

CamFlow extends the functionality of previous provenance systems, while retaining the same features.
Notably, we introduce a new completeness  trade-off against performance, storage, and I/O bandwidth.
In other words, we allow users to decide between capturing the entirety of supported system events (as PASS or LPM did) and capturing a subset of those events based of some particular well defined needs (see \autoref{sec:tailoring}).
Further, this same feature has proved helpful to new users when developing provenance applications (see \autoref{sec:examples}).
The tailoring feature has been used to help explain system behavior
and to help new developers get started.
A typical workflow with CamFlow is: 1) start capturing a single or small group of processes and their interactions; 2) develop a small prototype provenance application; and 3) move to an architecture able to handle scale issues introduced by whole-system provenance.
This has proved extremely beneficial as the accumulation rate and amount of provenance data can be daunting.
We and our users believe this to be important for increasing provenance adoption.

A remaining challenge is to provide a full comparison of CamFlow against alternative implementations: i.e., \emph{is the graph as complete and expressive as when using an alternative approach?}
To our knowledge, no study of such a nature has been completed.
Our experience tells us that CamFlow captures as much or more information compared to PASS or LPM, but we have not performed any formal evaluation. 
Indeed, quantitatively measuring provenance completeness or expressivity are complex, ongoing topics of research in the provenance community~\cite{chan2017expressiveness}.
Chan \etal~\cite{chan2017expressiveness} propose running a system-call benchmark and comparing the ``quality'' of the provenance captured.
However, one could argue that ``quality'' is a hard metric to evaluate.
An alternative approach is more formal and, as discussed in \autoref{sec:kernel:capture}, relies on the guarantees provided by LSM hooks that are
present on the path between any kernel object and a process.
However, recent research has demonstrated that this coverage may not capture every information flow~\cite{georget2017verifying}, as the LSM framework was originally conceived to implement MAC policy.
We derived from this publication~\cite{georget2017verifying} a patch to the LSM infrastructure~\cite{thomas_pasquier_2017_826436}.
This patch is used by most recent versions of CamFlow (>=0.3.3).
In addition to demonstrating that provenance is generated for every event, it must also be shown that the generated graph is ``correct'' and ``well connected''.
An empirical or formal demonstration of the completeness of the provenance captured is beyond the scope of this paper.

\section{Example Applications} \label{sec:examples}
We proposed that PaaS providers offer provenance as a service to their tenants.
In \autoref{sec:model}, we discussed how our model expands on previous provenance capture implementations, notably by recording information relevant to Docker containers.
In \autoref{sec:capture}, we showed how provenance can be captured and streamed in OS or distributed systems.
In \autoref{sec:tailoring}, we showed how the captured provenance can be tailored to meet the tenant application requirements.
Finally, in \autoref{sec:evaluation} we showed that this can be achieved with tolerable performance impact.
We now discuss, through examples, how cloud tenants can subscribe to the provenance data stream and develop provenance applications. We explore four such applications: two from from ongoing work at Harvard (\autoref{sec:examples:compliance} and \autoref{sec:examples:intrusion}),
and two where we implement examples from prior work, taking advantage of CamFlow's unique features (\autoref{sec:examples:DLP} and \autoref{sec:examples:server}).

\subsection{Demonstrable Compliance}
\label{sec:examples:compliance}

The manipulation of personal data in widely distributed systems, especially the cloud, has become a subject of public concern.
As a result, legislators around the world are passing laws to address issues around manipulation and privacy of personal data.
In Europe, data protection laws~\cite{eu-dpd, eu-gdpr} regulate and control all flows of personal data, in essence, information identifiable to individuals.
These flows must be for specific legitimate purposes, with various safeguards for individuals and responsibilities on those who hold, manage and operate on personal data.
In other jurisdictions, most notably the United States, which does not have such omnibus data protection laws, there are sector-specific restrictions on personal data. These include areas such as health~\cite{us-hipaa} and finance~\cite{us-sox}, as well as general Fair Information Practice Principles (FIPP)~\cite{federal2007fair},
which embody principles such as notice, choice, access, accuracy, data minimization, security, and accountability.
Further, there is pressure for the involved actors to provide transparency in how they handle data and to demonstrate compliance with the requirements.

In recent work~\cite{thomas_pasquier_2017_571433}, we developed an application that used CamFlow provenance to continuously monitor compliance with such regulations.
information flow constraints that translate into properties of paths in a provenance graph can express many of these regulations.
An \emph{auditor} can subscribe to a (distributed) system's provenance stream (as described in \autoref{sec:capture}) and verify that the stream respects the specified constraints, using a label propagation mechanism.
Developers write these constraints using
a framework similar to familiar graph processing tools such as GraphChi~\cite{kyrola2012graphchi} or GraphX~\cite{gonzalez2014graphx}.
The example application uses
structural and semantic properties of provenance graphs to reduce overhead,
claiming CPU overhead of about 5\% and storage of only a few hundred megabytes
of memory.
In parallel, the system retains the complete provenance record as forensic evidence, using cloud-scale storage such as Apache Accumulo~\cite{Moyer16}.
When the \emph{auditor} detects a regulation violation, users can query and analyze the forensic provenance
to identify the cause of the violation and take action, for example, by fixing
the system or notifying affected users as mandated by HIPAA~\cite{us-hipaa}.

\subsection{Intrusion Detection}
\label{sec:examples:intrusion}

The FRAPpuccino fault/intrusion detection system uses the capture mechanism presented here to identify errant or malicious instances of a PaaS application~\cite{han2017, han_2017_571444}.
An earlier system, pH~\cite{somayaji2002operating, somayaji2000automated}, recorded system call patterns of security-sensitive applications to detect abnormal behavior. We hypothesised that provenance data would be a better source of information, leading to more accurate results.
We experimented with reported real world vulnerabilities, and preliminary results indicate that we are able to detect attacks
with higher accuracy.

The system consists of two stages of operation: 1) a training stage where we build a per-application behavior model and 2) a deployment stage where we monitor deployed applications.
During the first stage, many instances of the target application run in parallel on the PaaS platform, and we assume most of them behave correctly.
We use the provenance graph generated to build a model of normal behavior.
In deployment, we analyze the provenance stream in sliding windows, comparing the current execution to the model generated during the training stage.
If the stream does not fit the model, we determine if there is a real intrusion or fault.
In the case of a false positive, we retrain the model incorporating the new data.
In the case of a true positive, we can analyze the forensic provenance to identify the root cause.

The prototype successfully detected known vulnerabilities, such as a \texttt{wget} arbitrary file upload~\cite{Golunski2016wget} that can be leveraged for privilege escalation, a tcpdump crash~\cite{Silveiro2016tcpdump} through malicious packets, and an out of memory exception in a Ruby server that breaks garbage collection~\cite{Gaziev2015ruby}.

This is ongoing work.
We plan to analyze streamed provenance data on Data-Intensive Scalable Computing systems such as Apache Spark, to expedite the process and make our intrusion detection system scalable. 
We will also extend the prototype to include the retraining process, to further improve accuracy in monitoring complex applications.
Work remains to evaluate the effectiveness of the approach 
in discovering new vulnerabilities.

\subsection{Data Loss Prevention}
\label{sec:examples:DLP}

Bates \etal describe how to leverage whole-system provenance capture to build Provenance-Based Data Loss Prevention (PB-DLP)~\cite{bates2015trustworthy}.
PB-DLP issues a provenance query that follows well-defined edge types to determine data dependencies
on data before it leaves the system.
If the dependencies include any sensitive information, PB-DLP aborts the transfer.
Bates \etal acknowledge that the provenance graph grows without bound,
creating increasingly long query times.

Our first approach to this use case leveraged the knowledge that policy enforcement queries need only consider a set of well-defined edge types.
As query time is dependent on the overall graph size, 
we tailor the capture policy, restricting capture to only those edges of the specified type(s).
This significantly reduces the volume of data collected.

The second approach uses CamFlow's taint mechanism, introduced in \autoref{sec:tailoring}.
Using taint propagation, we can replace a potentially costly graph query with a much less expensive
taint check when data is about to leave the system.
Propagation can be tuned taint to apply only to specific types of node and/or edge.
As a result, we can
produce the same behavior produced by the (expensive) query of Bates \etal~\cite{bates2015trustworthy} with a time-constant check.
Further, we note that taint tracking has been proved effective in similar scenarios~\cite{enck2014taintdroid}.
In conjunction with tracking taint using this method, we can also choose to store either whole or
selective provenance as forensic evidence.
This allows the use of ``post-mortem'' queries to explore the chain of events leading to the disclosure attempt.
This approach reduces computational overhead, while simultaneously providing forensic evidence.

\subsection{Retrofitting existing applications}
\label{sec:examples:server}

Bates \etal also retrofitted provenance into 
a classic three-tier web application~\cite{bates2016retrofitting}.
They use LPM to capture whole-system provenance
and modify the web server to relate HTML and database queries.
A proxy between the web server and the database collects SQL queries,
which are transformed into application-level provenance for the database.
It is therefore possible to link an HTML response to database entries.
Before responding to a query, the system issues a provenance query on the combined
application and system data to check for leakage of sensitive data, which it then prevents.
This architecture avoids having to modify the database, but still requires modifications to
the server and adds an additional component (the proxy) into the system.
With CamFlow we can achieve the same result without modifying the existing three-tier web application, as follows:\\
1) We track (and optionally propagate tracking from) both server and database executables.
2) We configure the server's
\texttt{ErrorLog} as a provenance log file, as described in \autoref{sec:application} and set the \texttt{LogLevel} to collect sufficient information to relate an HTML request to an SQL query.
3) We capture network packets (\eg by adding the line \texttt{record=DB\_IP/32:DB\_PORT} in the capture policy), which allows us to capture the SQL queries, without using an intermediate proxy.

\section{Related Work} \label{sec:rw}
Most provenance capture systems work at the application level, e.g.~\cite{angelino2010starflow, macko2012general, lerner2014rdatatracker, murta2014noworkflow} or at the workflow level~\cite{davidson2007provenance, bowers2012declarative}.
These systems provide an incomplete record of what happened, which renders
some use cases impossible (see \autoref{sec:application}).
System level provenance provides information about the interactions \emph{between} programs, which is the key to supporting more complex use cases.
One of the first attempts at system-level provenance is the Lineage File System~\cite{Can2005} that used system call interposition in a modified Linux Kernel, while a user space process read captured provenance out of a \textsf{printk} buffer and stored the data in a SQL database.
PASS captures provenance at the Virtual File System (VFS) layers. PASSv1~\cite{muniswamy2006provenance} provides functionality to capture processes' I/O interactions, while PASSv2~\cite{muniswamy2009layering} introduced a disclosed provenance API to allow the integration of provenance across semantic levels.
The main issue for these systems was the difficulty of maintaining and extending provenance capture over time, since it was spread among various places in the kernel.
CamFlow is a strictly self-contained implementation that uses standard kernel features and is extremely easy to extend and maintain.

The next generation of system-level provenance used the Linux Security Framework (LSM) to capture provenance.
The LSM framework provides a guarantee of completeness~\cite{edwards2002runtime, jaeger2004consistency, ganapathy2005automatic} of the provenance captured;
Hi-Fi~\cite{pohly2012hi} was the first such approach.
Hi-Fi did not support security module stacking (\ie SELinux or AppArmor cannot run alongside provenance capture), rendering the system it monitors vulnerable.
Further, Hi-Fi did not provide any API for disclosed provenance.
Linux Provenance Module (LPM)~\cite{bates2015trustworthy} addressed those limitations.
LPM provides an API for disclosed provenance and duplicates the LSM framework to allow the stacking of provenance and MAC policy.

CamFlow leverages recent advances of the LSM framework~\cite{edge2015, Schaufler2016} to allow stacking of LSMs without feature duplication, thereby further increasing maintainability and extensibility.
A second objective of LPM was to provide stronger integrity guarantees: leveraging TPM and the Linux Integrity Measurement Architecture~\cite{jaeger2006prima, sailer2004design} and using cryptographic techniques to protect packet-level provenance taint.
We can do the former, while the latter tends to create compatibility issues.
We believe in a clearer separation of concerns, leaving packet integrity guarantees to mechanisms such as IPSec~\cite{frankel2011ip}.

An alternative implementation of whole-system provenance is
interposition between system calls and libraries in user space, as in OPUS~\cite{balakrishnan2013opus}.
An argument in favour of such systems is that modifications to existing libraries are more likely to be adopted than modifications to the kernel. 
However, for this approach to work, \textbf{all} applications in the system need to be built against, or dynamically linked to, provenance-aware libraries, replacing existing libraries.
Further, the guarantees of completeness provided are much weaker than those provided by an LSM-based implementation.
SPADE~\cite{gehani2012spade} mixes this approach, with the Linux Audit infractustructure; the LSM-based approach to provenance provides a more detailed record and stronger completeness guarantees.

Another system-call based approach that uses both logging and taint tracking is ProTracer~\cite{shiqing2016}.
ProTracer relies on kernel tracepoints to intercept system calls and capture provenance.
The system relies on taint propagation when the events being traced are
ephemeral (\eg inter-process communication) and full provenance capture for events that are determined to be permanent.
One concern with system call-based approaches is the lack of guarantees of completeness in the provenance record.
Consider for example the {\tt read} and {\tt pread64} system calls.
Both read from a file descriptor, the difference being that the latter can do so with an offset.
A system needs to properly record both types of read system calls, and should probably
ensure that the record for {\tt pread64} is the same as the record of an {\tt lseek} and subsequent {\tt read} call.
Ensuring such consistency increases the complexity of the capture (or post-processing).
LSM-based approaches are guaranteed to capture every event that is deemed security-sensitive and focus on the objects being accessed, instead of the actions being carried out on those objects.

One of the main hurdles of system-level provenance capture is the sheer amount of data generated~\cite{bates2015trustworthy, Moyer16}.
One approach to this issue is to improve provenance ingest to
storage~\cite{Moyer16} and provenance query~\cite{vicknair2010comparison}.
A second approach is to reduce the amount of provenance data captured.
This reduction might be based on security policies~\cite{bates2015take, pasquier2016ic2e},
limiting capture to
only those elements considered \emph{important} or \emph{sensitive}.
This is of limited interest in contexts such as scientific workflows, where data of interest may not be associated with any security policy.
We address these issues by providing administrators and system designers with selective capture
in a way consistent with the provenance data-model.

Using system-level provenance in a cloud computing context has been proposed in the past~\cite{muniswamy2009making, muniswamy2010provenance, zhang2011track, lee2015towards}. To our knowledge, the work presented here is the first open-source maintained implementation with demonstrated use cases (\autoref{sec:examples}).
The main focus of our work has been to develop a practical system that can be used and expanded upon by the community.
There is application-level provenance for a cloud computational framework such as MapReduce~\cite{akoush2013hadoopprov} or Spark~\cite{interlandi2015titian, gulzar2016interactive}. We see these as complementary; they could be incorporated in  CamFlow provenance capture through the mechanism discussed in \autoref{sec:kernel:interface}.

\section{Conclusion} \label{sec:conclusion}
CamFlow is a flexible, efficient and easy to use provenance capture mechanism for Linux. Linux is widely used in cloud service offerings and our modular architecture, using LSM, means that data provenance can be included without difficulty as part of a cloud OS in PaaS clouds.

We take advantage of developments in Linux kernel architecture to create a modular, easily maintainable and deployable provenance capture mechanism. 
The ability to finely tune the provenance
being captured to meet each application's requirements helps to create a scalable system.
Our standards-compliant approach further enhances the broad applicability and ease of adoption of CamFlow.
Further, we demonstrated the low impact on performance and the wide applicability of our approach.

\section*{Acknowledgments}
\noindent This work was supported by the US National Science Foundation under grant SSI-1450277 End-to-End Provenance. 
Early versions of CamFlow open source software were supported by
UK Engineering and Physical Sciences Research Council grant EP/K011510 CloudSafetyNet.

\bibliographystyle{ACM-Reference-Format}
\bibliography{biblio}

\appendix

\section{Availability}
\label{sec:availability}
The work presented in this paper is open-source and available for download at \url{http://camflow.org} under a GPL-3.0 license. 
Detailed installation instructions are available online.\endnote{\url{https://github.com/CamFlow/CamFlow.github.io/wiki/Getting-Started}}

 \subsection{Running CamFlow}

To get started with the system presented in this paper install Vagrant\endnote{\url{https://www.vagrantup.com/}} and VirtualBox\endnote{\url{https://www.virtualbox.org/}} then run the following commands:
\begin{lstlisting}[language=bash, style=mystyle, caption={Running CamFlow demo in Vagrant.}, label=listing:availability:install]
git clone https://github.com/CamFlow/vagrant.git
cd ./vagrant/basic-fedora
vagrant plugin install vagrant-vbguest
vagrant up
# testing the installation
vagrant ssh
# check installed version against CamFlow head
camflow -v
uname -r
# check services
cat /tmp/audit.log # audit service logs
cat /tmp/camflow.clg # configuration logs
\end{lstlisting}

Alternatively, CamFlow can be installed and kept up to date via the package manager.
The packages are distributed via packagecloud\endnote{\url{https://packagecloud.io/camflow}} and available for Fedora distributions.
Installation on a Fedora machine, do as follow:
\begin{lstlisting}[language=bash, style=mystyle, caption={Installing CamFlow on Fedora.}, label=listing:availability:package]
curl -s https://packagecloud.io/install/repositories/camflow/provenance/script.rpm.sh.rpm | sudo bash
sudo dnf install camflow
\end{lstlisting}

\subsection{Repeating Evaluation Results}

The instructions for reproducing the results presented in \autoref{sec:evaluation} are online.\endnote{\url{https://github.com/CamFlow/benchmark}}
We assume that a Linux kernel running CamFlow has been installed. To run the various tests please follow the instructions:

\begin{lstlisting}[language=bash, style=mystyle, caption={Reproducing evaluation results.}, label=listing:availability:reproduce]
git clone https://github.com/CamFlow/benchmark.git
cd benchmark
make prepare
# choose the config to run the tests under
make <whole/selective/off>
make run
# for more accurate results you may want
# to run test individually and reboot
# after each. Run for example:
make run_lmbench
\end{lstlisting}
 
\theendnotes
\end{document}